\documentclass[useAMS,usenatbib]{mn2e}
\usepackage{graphicx,epsfig}

\def\asec{\ifmmode ^{\prime\prime}\else$^{\prime\prime}$\fi}

\def\msun{\hbox{~M$_{\odot}$}}

\def\lsun{\hbox{~L$_{\odot}$}}
\def\msunyr{\mbox{\,${\rm M_{\odot}\, yr^{-1}}$}}
\def\mdot{~\dot M}

\def\degs{\ifmmode ^{\circ}\else$^{\circ}$\fi}
\def\amin{\ifmmode ^{\prime}\else$^{\prime}$\fi}
\def\asec{\ifmmode ^{\prime\prime}\else$^{\prime\prime}$\fi}

\def\degs{\ifmmode ^{\circ}\else$^{\circ}$\fi}
\def\amin{\ifmmode ^{\prime}\else$^{\prime}$\fi}

\def\EE#1{\times 10^{#1}}

\def\cm{\mbox{\,cm}}

\def\cm3{\mbox{\,cm$^{-3}$}}

\def\lsim{\!\!\!\phantom{\le}\smash{\buildrel{}\over
 {\lower2.5dd\hbox{$\buildrel{\lower2dd\hbox{$\displaystyle<$}}\over
                                 \sim$}}}\,\,}
\def\gsim{\!\!\!\phantom{\ge}\smash{\buildrel{}\over
{\lower2.5dd\hbox{$\buildrel{\lower2dd\hbox{$\displaystyle>$}}\over
                               \sim$}}}\,\,}
\def\lfir{L_{\rm FIR}}
\def\lsyn{L_{\rm syn}}

\def\whz{~W~Hz$^{-1}$}
\def\snrate{\nu_{\rm SN}}

\title{A Supernova Factory in Mrk 273?}
\author[M. Bondi et al.]
       {M.~Bondi,$^1$ M-A. P\'erez-Torres,$^2$ D. Dallacasa,$^{3}$
        T.W.B. Muxlow$^{4}$ \\
        $^1$INAF-Istituto di Radioastronomia, Via Gobetti 101, I-40129, Bologna,
        Italy\\
        $^2$Instituto de Astrof\'{\i}sica de Andaluc\'{\i}a, CSIC, Apartado Correos
	3004, 18080 Granada, Spain\\
        $^3$Dipartimento di Astronomia, Universit\`a di Bologna, Via Ranzani 1,
        I-40127 Bologna, Italy\\
	$^4$Jodrell Bank Observatory, University of Manchester, Macclesfield,
	Cheshire, SK119DL, U.K.\\
}
\begin{document}
\maketitle

\begin{abstract}
We report on 1.6 and 5.0 GHz observations of the ultraluminous infrared 
galaxy (ULIRG) Mrk~273, using the European VLBI Network (EVN) and the
Multi-Element Radio-Linked Interferometer Network (MERLIN). 
We also make use of published 1.4 GHz VLBA observations of Mrk~273 by 
Carilli \& Taylor (2000)\nocite{CT00}.
Our 5 GHz images have a maximum resolution of 5--10 mas, which corresponds
to linear resolutions of 3.5--7 pc at the distance of Mrk~273, and are
the most sensitive high-resolution radio observations yet made of this 
ULIRG.  
Component N1, often pinpointed as a possible AGN, displays a steep 
spectral index ($\alpha = 1.2 \pm 0.1; S_\nu\, \propto\, \nu^{-\alpha}$); 
hence it is very difficult to reconcile with N1 being an AGN, 
and rather suggests that the compact nonthermal radio emission is produced
by an extremely high luminous individual radio supernova (RSN),  or a
combination of unresolved emission from nested supernova remnants (SNR),  
luminous RSNe, or both.
Component N2 is partly resolved out into several compact radio 
sources --none of which clearly dominates-- and
a region of extended emission about 30 pc in size.
The integrated spectral index of this region is flat 
($\alpha = 0.15 \pm 0.1$), which can be interpreted as due
to a superposition of several unresolved components, e.g., 
RSNe or SNRs, whose radio emission peaks
at different frequencies and is partially free-free absorbed.
Is it also possible that one of the compact components detected in this 
region is the radio counterpart of the AGN.
The overall extended radio emission from component N is typical of
nonthermal, optically thin radio emission
($\alpha = 0.8 \pm 0.1$), and its 1.4 GHz luminosity 
($L_{1.4 \rm GHz} = (2.2 \pm 0.1)\times 10^{23} $ WHz$^{-1}$) 
is consistent with being produced by 
relativistic electrons diffused away from supernova remnants in an outburst.
The southern component, SE, shows also a very steep spectrum
($\alpha = 1.4 \pm 0.2$), and extended radio emission whose origin and
physical interpretation is not straightforward.
\end{abstract}


\section{Introduction}
\label{Intro}
Ultraluminous infrared galaxies (ULIRGs) are the most luminous galaxies
in the local universe, with luminosities exceeding $10^{12}$ L$_\odot$
\cite{Sand88}.
ULIRGs are associated with merging systems, where large quantities of gas and
dust are channeled in the inner regions and heated by a powerful source of
optical-uv continuum. 
The most popular dust-heating mechanisms in ULIRGs propose the 
existence of either an active galactic nucleus (AGN), or a massive starburst, 
but which one is the dominant mechanism is hitherto an open question.
Recent near-IR spectroscopic measurements suggest enhanced star
formation in the majority of ULIRGs \cite{Genz98}, with a
significant heating from the AGN only in the most luminous objects 
(Veilleux, Sanders \& Kim 1999)\nocite{VSK99}.

\begin{figure*}
\includegraphics[width=5.5cm, angle=-90]{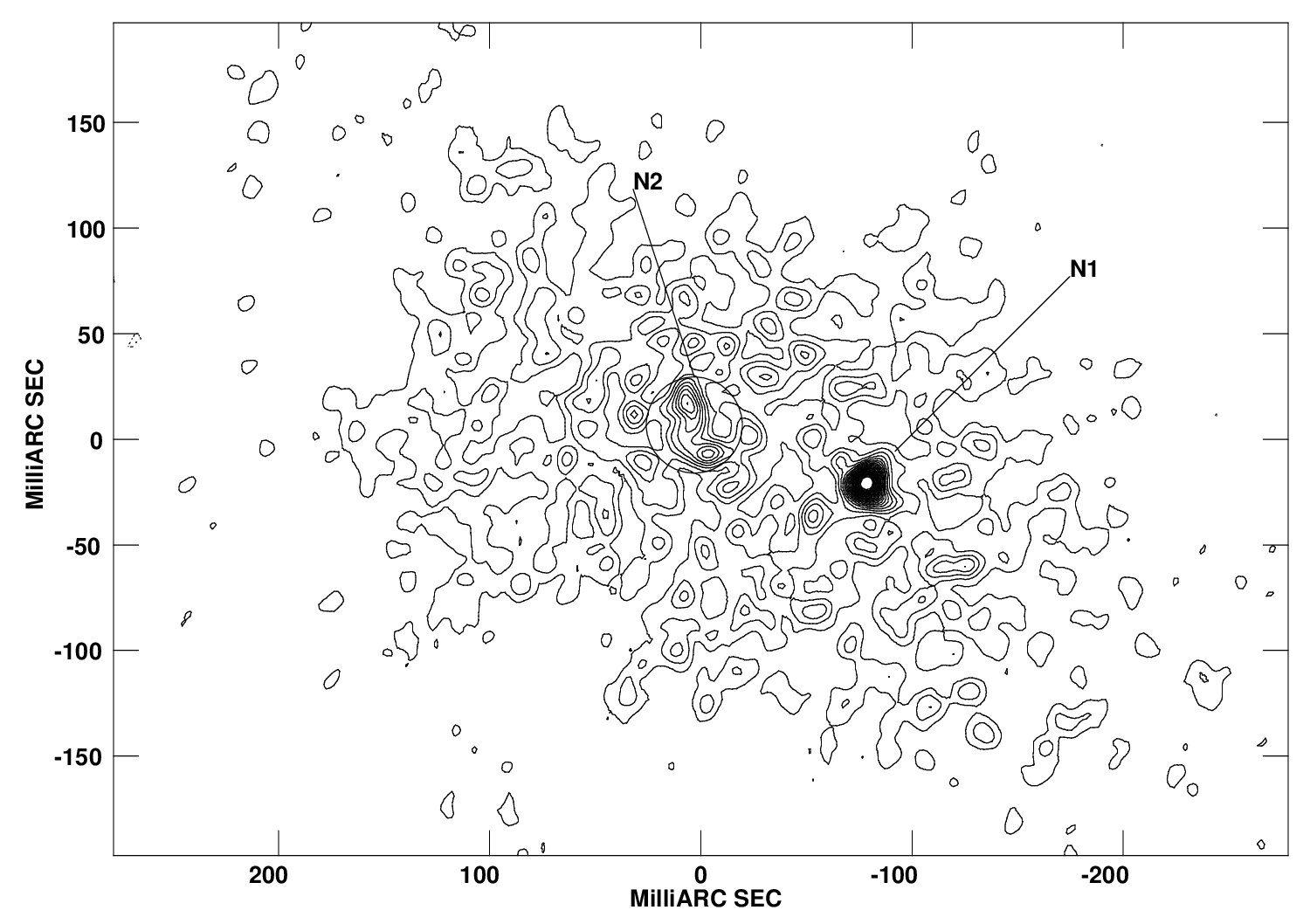}
\includegraphics[width=5.5cm, angle=-90]{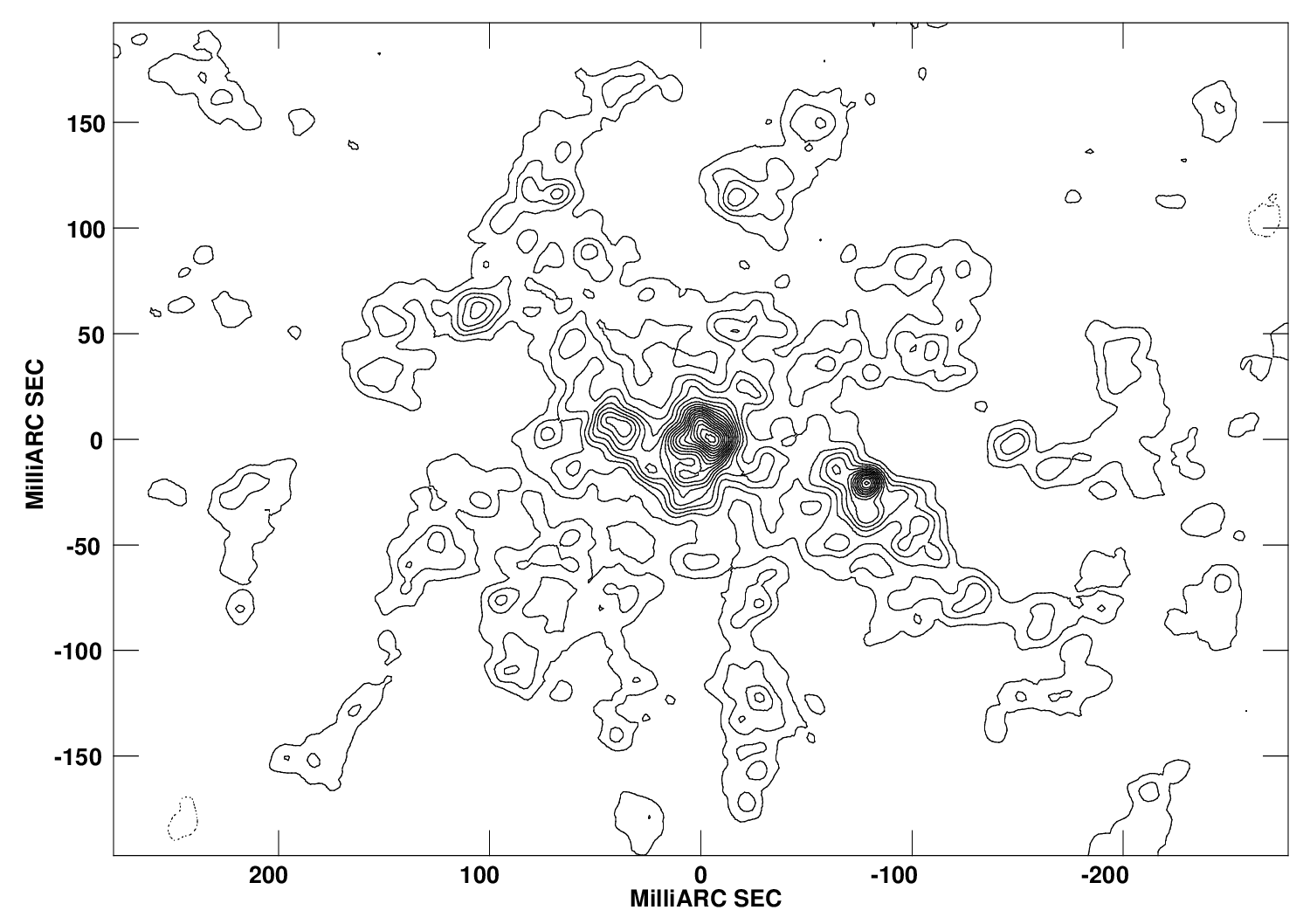}
\caption{
{\it a) Left:} Image of component N of Mrk 273 at 1.4 GHz at 10 mas
resolution from Carilli \& Taylor 2000.
The contours are linear with an increment of 0.1 mJy/beam,
starting at 0.1 mJy/beam. The peak surface brightness is 3.05 mJy/beam, and
the off-source rms is 36 $\mu$Jy/beam. Labels point to components discussed
in the text. {\it b) Right:} Component N of Mrk 273
at 5 GHz, 10 mas resolution. The contours are $-1, 1, 2, 3, 4, {\ldots}
\times 0.04$ mJy/beam. The peak surface brightness is 0.74 mJy/beam and the
off-source noise is 15 $\mu$Jy/beam.
}
\end{figure*}

Radio observations can prove to be extremely useful in shedding light to 
the relevant question of the dust-heating mechanism in ULIRGs, 
as they are unaffected by dust extinction and allow for sub-parsec 
resolution, using VLBI techniques.
The most spectacular evidence to date of a dominant starburst in a ULIRG is
the discovery of a population of bright radio supernovae (RSNe) in the nuclear
region of Arp~220, detected at 1.6 GHz \cite{Smit98}.

Mrk~273 is a ULIRG at $z=0.0378$ ($D\simeq 150$ Mpc for $H_0=70$
kms$^{-1}$Mpc$^{-1}$) classified as a Seyfert 2 and/or LINER
merging system showing a disturbed morphology on the kpc scale.
The nuclear region of Mrk~273 is extremely complex and has been
studied in detail in the radio (Cole et al. 1999\nocite{Cole99};
Carilli \& Taylor 2000\nocite{CT00}, hereafter CT00; Yates et al.
2000\nocite{Yate00}), near infrared \cite{Knap97}, optical \cite{MB93},
and X-rays \cite{Xia02}.

In this paper, we present high-resolution imaging (5-10 mas) 
of the 1.6 and 5 GHz continuum radio emission of Mrk~273, obtained
with the European VLBI Network (EVN) and the Multi-Element Radio-Linked
Interferometer Network (MERLIN). These data confirm the presence of an
extremely strong starburst in component N, which we start resolving out in 
a large number of compact components, suggestive of a probable supernova 
factory in this region. 
The presence of a possible AGN in the nuclear environment
of Mrk~273 is still elusive: its identification with
component N1 is highly unlikely because of its radio spectrum;  we also set up 
a stringent upper limit on the radio emission of an AGN nucleus in component
SE. Despite the high sensitivity and milliarcsecond resolution of our
observations, the nature of component SE is still unclear.

\section{Previous high resolution observations of the nuclear region in 
Mrk~273}
\label{Radio}

Mrk~273 shows three extended radio components 
(conventionally named N, SE and SW, see Fig. 1 in Yates et al. 
2000\nocite{Yate00}) within 1 arcsecond. These are physically
related components, probably associated with the merging process, and
not the result of chance projection of background sources.
The three components show different morphologies and properties.
Component SW is extended, of very low surface brightness and barely
detected in the radio, while components N and SE are much brighter.
In particular, component N shows two peaks, N1 and N2,
embedded in extended radio emission.
Component N
and SW have bright NIR counterparts (Majewski et al. 1993\nocite{Maje93};
Knapen et al. 1997\nocite{Knap97})
and are redder than component SE, suggesting the presence of strong star
formation, while component SE is not detected in the
NIR.

CT00 observed Mrk~273 in June 1999 at 1.4 GHz using the
Very Long Baseline Array (VLBA) and the full Very Large Array (VLA)
to image component N and SE with 10 mas resolution.
They resolved component N (Fig. 1{\it a}) in a number of
compact features embedded in a weak and diffuse radio emitting region, and
suggested that one of them, coincident with N1,
could be associated with a weak AGN, while the other
fainter compact features, mainly in a region around N2, could be nested SNR, 
luminous RSN, or both.
While there is evidence for the presence of a weak AGN in component N
(e.g., discovery of hard X--ray emission, Xia et al. 2002\nocite{Xia02};
OH megamaser emission, Kl\"ockner \& Baan 2004\nocite{KB04})
the AGN cannot possibly account for more than a few percent of the
observed radio emission.
\begin{figure*}
\includegraphics[width=5cm, angle=-90]{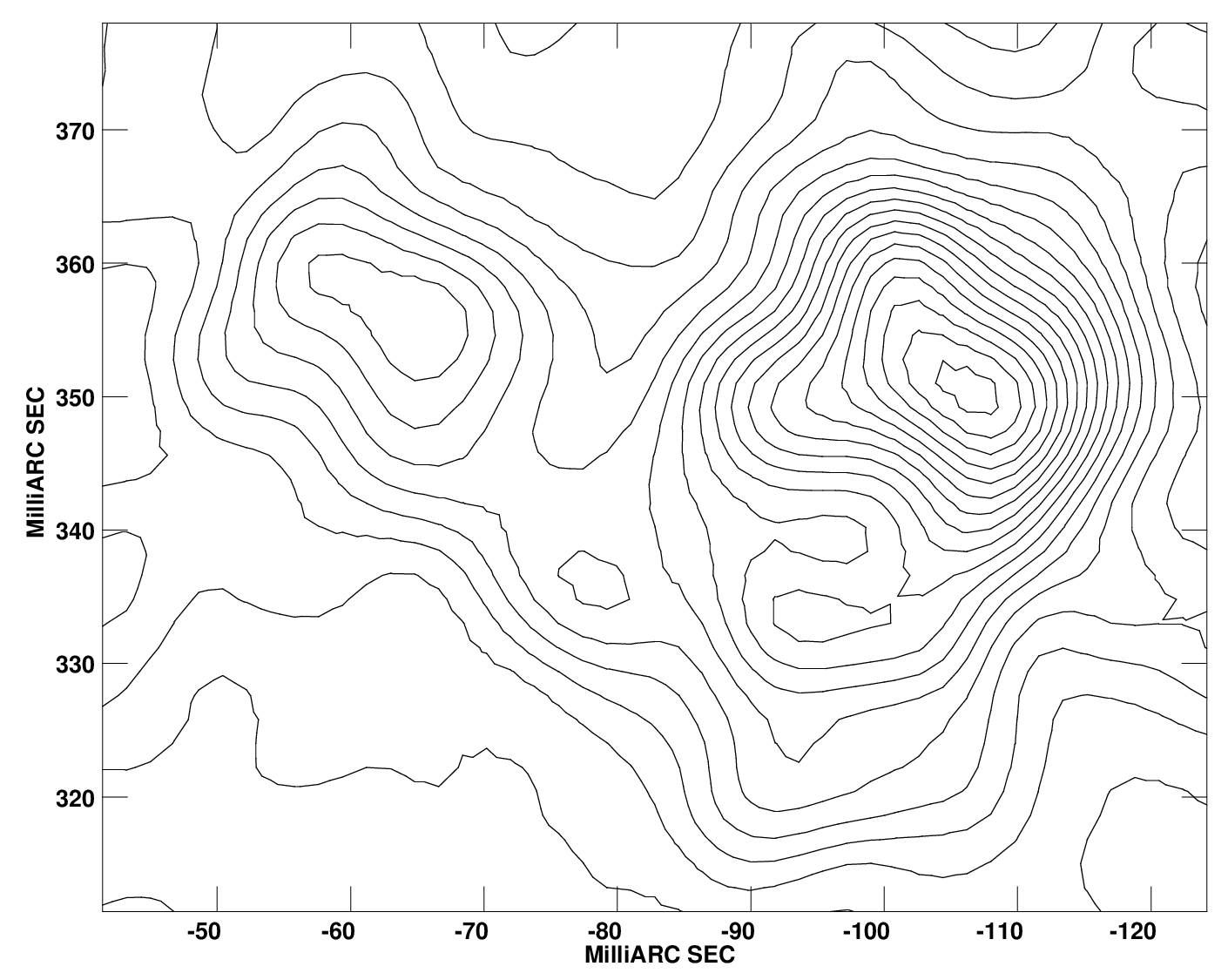}
\includegraphics[width=5cm, angle=-90]{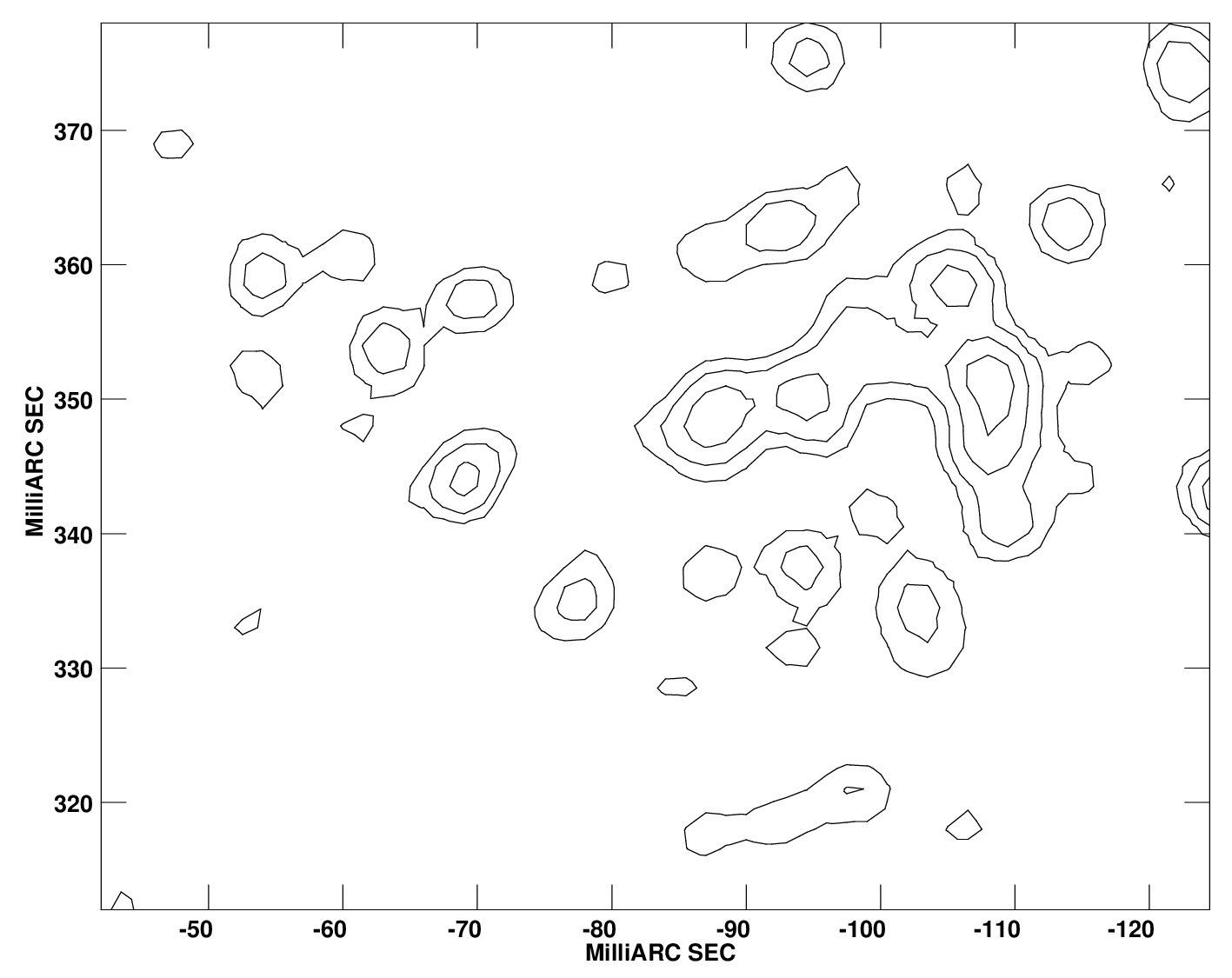}
\caption{Blow out of the N2 region from the 5 GHz EVN+MERLIN image at
10 mas resolution ({\it left}) and 5 mas resolution ({\it right}). First
countour in the 5 mas image is 0.1 mJy/beam.
}
\end{figure*}

The interpretation of the radio morphology of component SE is less
straightforward: it shows (Fig. 3{\it a}) an elongated structure of about 
40 mas long in 
the N-S direction,  embedded in a weak halo, which is consistent with an 
amorphous jet or a very compact starburst.
The lack of NIR emission in component SE could argue in favour of the AGN
interpretation for this source, but the possibility that this component
is still obscured at 2.2 $\mu$m cannot be ruled out.

\section{Radio Observations}
We observed Mrk~273 at 5~GHz using the EVN and
MERLIN and at 1.6 GHz (EVN-only) in February 2004,
with the main goal of performing a spectral analysis of the
compact and extended features in the N and SE components. 
The observations were carried at 512 Mbit/s
sustained bit rate to exploit the large bandwidth capabilities of the EVN,
with an array which included all the European antennas. These were the first
observations at 5 GHz for the resurfaced Lovell telescope.
Mrk~273 was observed in phase-reference mode for a total on-source time of
5.5 hours at both frequencies.
The compact source J1337+550 was observed every 5 minutes as phase reference,
while OQ208 and J1310+322 were used to calibrate the bandpass.
Data reduction was performed using the Astronomical Image Processing System
(AIPS). Standard a priori gain calibration was performed using the measured
gains and system temperatures of each antenna.
The amplitude calibration was refined using the phase reference source.
While the 5 GHz observations were of very high quality,
the 1.6 GHz data suffered from failures in some of the large antennas, which
prevented us from reaching the necessary accuracy and sensitivity to properly
image the complex and extended emission in Mrk~273.
For this reason, we resorted to use the 1.4 GHz observations obtained by CT00
in order to derive the spectral index properties of components N and SE.
The $1\sigma$ r.m.s. noise is 15 $\mu$Jy and 36 $\mu$Jy for the EVN+MERLIN
5 GHz and EVN-only 1.6 GHz observations, respectively.
Images at different frequencies were aligned using component N1.
\section{Discussion}
\label{Disc}

\subsection{The Northern Component}
The northern source in Mrk 273 consists of two regions of compact emission,
coincident with the N1 and N2 components of the lower resolution MERLIN image,
embedded in a halo of diffuse emission of roughly the same extension of that 
detected at 1.4 GHz by Carilli \& Taylor (see Fig. 1). 

At this resolution the extended emission is heavily resolved, and 
the CLEAN algorithm tends to generate spurious compact sources
when deconvolving diffuse emission regions. For this reason, a
point-by-point comparison between the 1.4 GHz and 5 GHz images in the
extended emission region is meaningless. 
On the other hand, the point-like component N1
and the region N2 (with a size of about 50 mas) are well defined and are 
likely not to be affected by deconvolution problems. 
It has been usually assumed in the
literature that N1 hosts a weak AGN nucleus, while N2 is a very compact region
of massive star formation. This view is supported by two major points:
1) the detection of high excitation IR lines \cite{Genz98}) and hard X-ray 
emission \cite{Xia02} are evidences of an AGN-like nucleus in Mrk 273; 
2) the CO emission and NIR peaks are spatially coincident
with N2 (Knapen et al. 1997\nocite{Knap97}; Downes \& Solomom 
1998\nocite{DS98}).

We find that component N1 
has a very steep spectral index ($\alpha\simeq 1.2; S \propto \nu^{-\alpha}$)
between 1.4 and 5 GHz, 
and a luminosity at 1.4 GHz of about $7\times 10^{21}$ WHz$^{-1}$.
The flux density of N1 derived from the lower quality 1.6 GHz observations
(at the same epoch of the 5 GHz data) is consistent with a steep spectral
index of about 1, so we can rule out that strong flux density variability
between the CT00 observations in 1999 and ours in 2004
is responsible for the steep spectral index of component N1.
While the radio luminosity falls in the range typical of radio nuclei in low
luminosity AGNs and Seyfert galaxies (e.g. Falcke et al. 2000\nocite{Falc00}; 
Ulvestad \& Ho 2001\nocite{UH01}), the radio spectrum of these nuclei is, 
in the great majority of cases, flat or even inverted (Nagar, Wilson \& Falcke
2001\nocite{NWF01}; Middelberg et al. 2004\nocite{Midd04}).
On the other hand,
the compact steep spectrum (CSS) and giga-Hertz peaked spectrum (GPS) sources
have  compact pc-scale morphology and
steep radio spectrum but are 3-4 orders of magnitude more luminous than N1
\cite{Odea98}. 
Such properties are difficult to reconcile with N1 being
an AGN, and rather suggests a supernova origin for the nonthermal radio 
emission.
The steep spectral index is indeed typical of a young RSN 
(e.g., Weiler et al. 2002\nocite{Weil02}),
but in this case it would be an extremely luminous one. In fact, the average
maximum luminosity of typical radio supernovae is around $2\EE{20}$ WHz$^{-1}$,
and the most luminous type IIn radio SN, like SN1988Z, are still a factor of 
three to four less luminous at maximum than N1. Given this premise, a realistic
scenario would be one in which N1 is not an individual radio SN, but a
combination of nested supernova remnants, several luminous radio supernovae, 
or both.
Higher resolution multifrequency observations are needed to clarify the
nature of component N1.

Component N2 shows a more complex morphology.
It occupies a well defined region of about 50 mas in size,
partly resolved into several compact radio sources, whose
integrated spectral index is flat ($\alpha \simeq 0.15$).
Such a flat spectrum could be due
to a superposition of several, unresolved components 
whose radio emission peaks at different frequencies and is partially 
free-free absorbed.
In Figure 2 we show a blow out of the N2 region, obtained with a resolution
of 10 mas and 5 mas. In the full resolution image, we resolve most of the
emission in N2 in compact features which we tentatively identify with RSN
and/or SNR. These compact features are weak (peak brightness
in the range 0.1-0.4 mJy/beam) and the possibility that some of them might
be spurious components, produced by resolving out the extended emission in N2,
cannot be excluded. On the other hand, we note that most of these compact
features can be identified in the 10 mas resolution image as well, where
the N2 region is not resolved out or affected by deconvolution problems.
In order to test the fidelity of the flux density measurements of the 
compact features in our images we ran the following test. We added, in the
{\it u-v} plane, 10 point-like components in the central region of
Mrk 273 N with random positions and peak flux densities in the range
0.1--0.4 mJy/beam using the task UVMOD in AIPS. We then cleaned the image in
the same manner as we did for the image shown in Fig. 2b, and derived the
peak flux of the components previously injected. We repeated this exercise
30 times, obtaining a sample of 300 simulated sources. 
For each source we derived the
difference between the measured and real flux normalized by the  r.m.s.
noise ($\sigma$). Only 14/300 ($4.6\%$) and 55/300 ($18\%$) sources have 
measured  fluxes which are different from the real values by more than 
$3\sigma$ and $2\sigma$, respectively. 
Considering we injected these sources in an area
where other components are already present, and that we did not make any
attempt to discriminate among these cases, the test reassures us about the
fidelity of the flux densities of the compact components derived from the
images.

The radio morphology of the region N2 suggests
the presence of multiple compact components in an area  of about 30
pc of size, and this is also consistent with the results from CO and NIR 
emission
observations which indicate this region as the core of an extremely rich
star forming region.
\begin{figure*}
\includegraphics[width=4.cm]{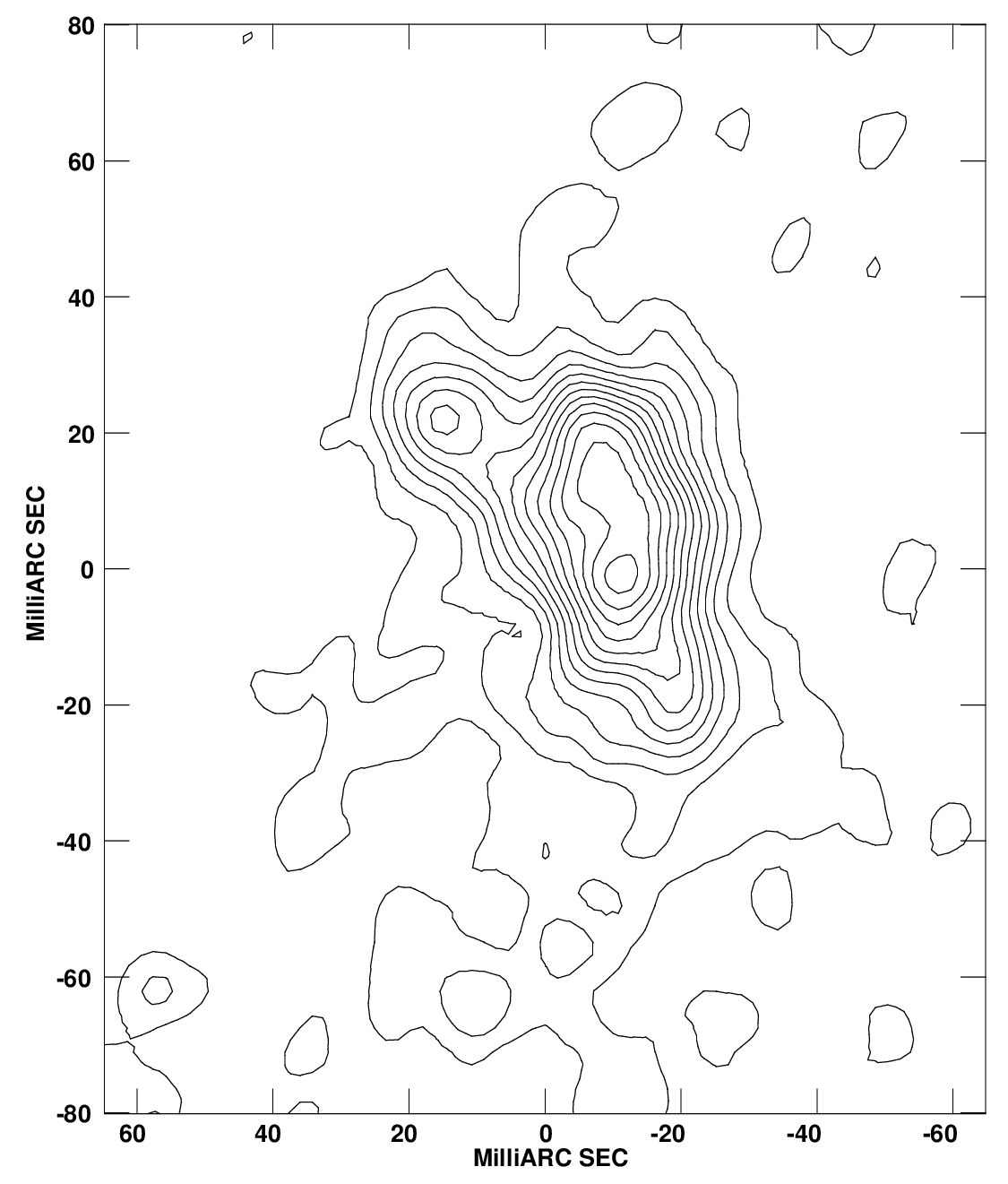}
\includegraphics[width=4.cm]{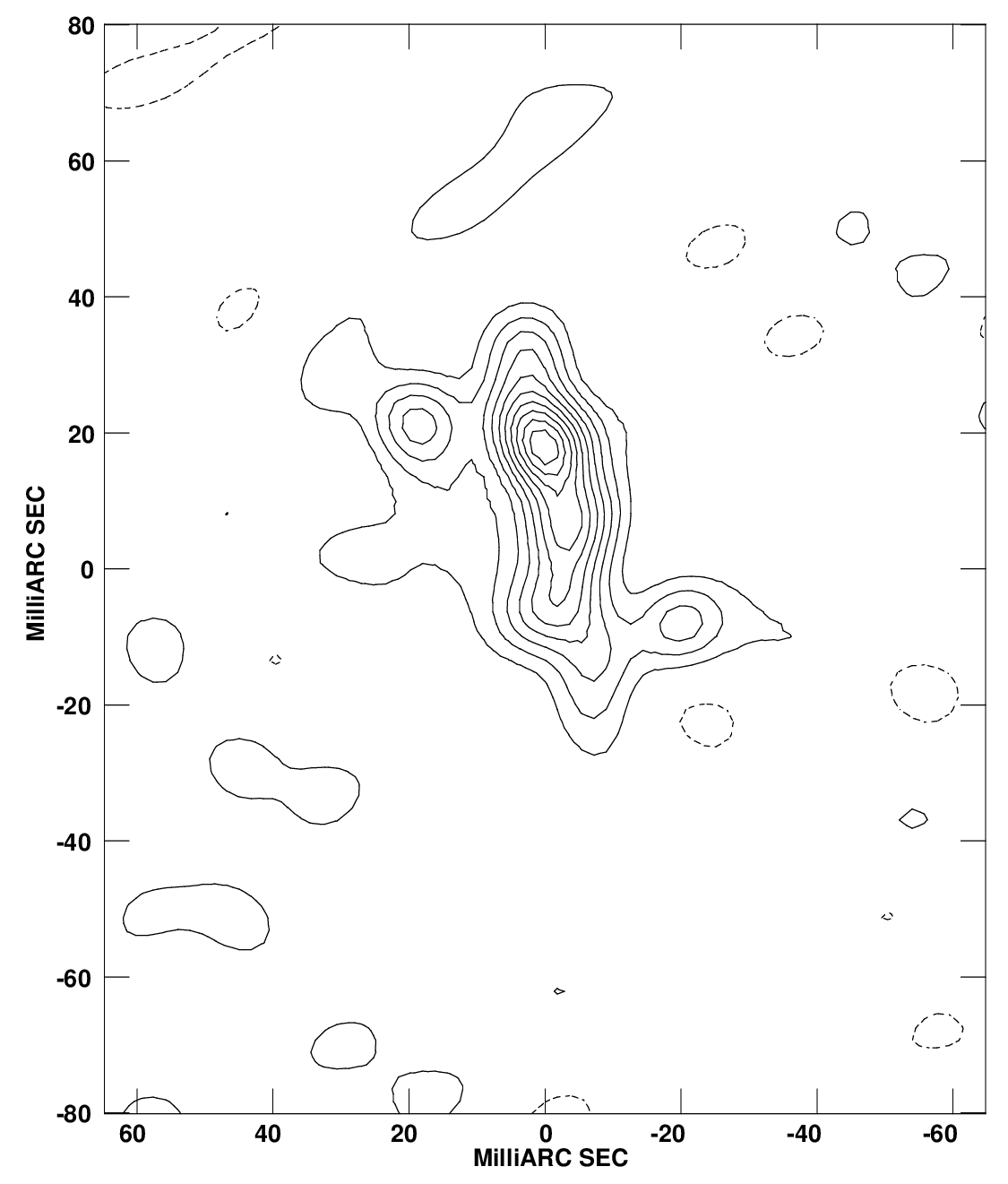}
\includegraphics[width=4.cm]{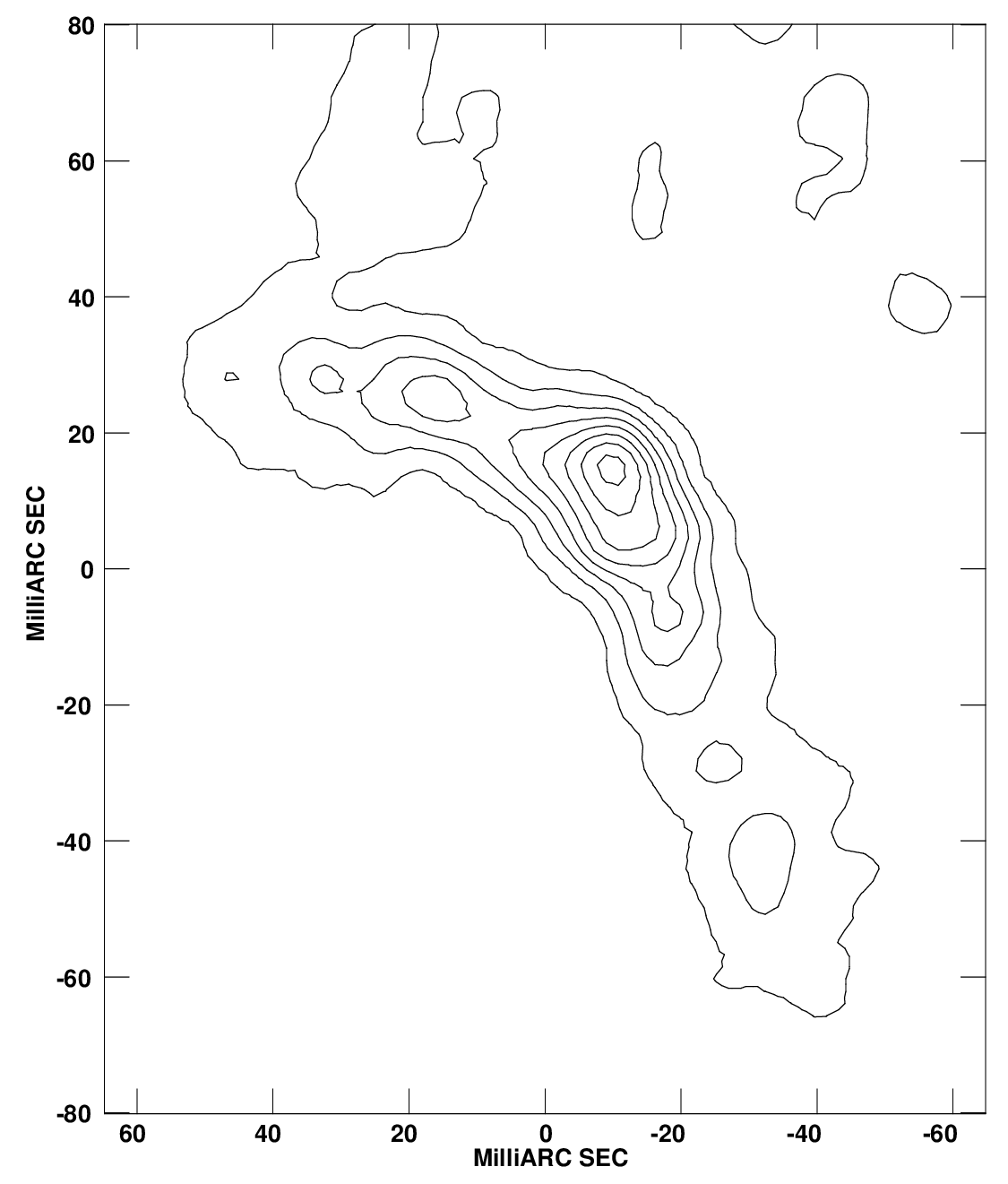}
\caption{Component SE in Mrk 273 imaged with a 10 mas beam. {\it a) Left:}
VLBA+VLA at 1.4 GHz (June 1999), contours and noise as fig. 1a; {\it b)
Middle:} EVN at 1.6 GHz (Feb 2004), first contour 0.11 mJy/beam linear
increment, noise 35 $\mu$Jy/beam.
{\it c) Right:} EVN+MERLIN at 5 GHz (Feb 2004), contours and noise as fig. 1b.
}
\label{fig,SE}
\end{figure*}

Downes \& Solomon (1998)\nocite{DS98} derived an IR luminosity of 
$6.0\times 10^{11} L_\odot$ generated in a  region of radius $\sim$120 pc
around components N1 and N2, and estimated 
a molecular mass of $\approx\,1.0\times 10^9\,M_\odot$, 
and a mass in new stars (i.e., excluding matter lost from the earliest
high-mass stars during their post-main-sequence evolution) 
of $M_\star \approx \,1.6\times 10^9\,M_\odot$. 
The above values are useful to constrain scenarios in which 
the luminosity in the nuclear region of Mrk~273 arises from a starburst.
We follow here the prescription by Scoville, Yun \& Bryant (1997)
\nocite{SYB97}, 
which characterize a starburst by its total luminosity, $L$, Ly continuum 
production, $Q$, and accumulated stellar mass, $M_\star$. 
Those values are obtained as power-law approximations of the lower
and upper mass cutoffs for stars, $m_l$ and $m_u$, 
the constant rate of star formation, $\mdot$, and the burst timescale, $t_B$.
In particular, for a burst timescale of $t_B = 5\EE{7}$\,yr 
a simultaneous match for the values of $\lfir$ and $M_\star$
yields $m_l \approx 3.0 \msun$, $m_u \approx 30\msun$, and $\mdot \approx 39 \msunyr$.
The implied ionizing photon rate is 
$Q \approx 3.9\EE{44}$s$^{-1}$, about a factor of two smaller than for Arp~220.
Here, we used a minimum mass for yielding type II supernovave of 8$\msun$, 
which results in a supernova rate $\snrate \approx 1.3$\,yr$^{-1}$.  
(For longer burst timescales, e.g., $t_B = 10^8$~yr, the upper mass limit
can be increased to about 35$\msun$, while $\mdot$
can be decreased to $\approx 31\msun$, implying a supernova rate of 
$\approx 1.0$\,yr$^{-1}$.) 
Starbursts with very low-mass stars ($m_l = 0.1 \msun$) can probably 
be ruled out, as they require a mass in stars about an order of magnitude 
greater than expected. 

We also used our radio flux density measurements of component N to constrain
the range of plausible models for a starburst in Mrk~273. 
The extended radio emission of the northern component has a total 
flux of $\sim 80$ mJy at 1.4 GHz, and an average 
spectral index between 1.4 and 5 GHz of $\alpha\approx 0.8$,
yielding an observed luminosity $L\simeq 2.2\times 10^{23}$~W~Hz$^{-1}$.
Condon (1992)\nocite{Cond92} gives a number of simple scaling law relationship to
estimate starburst characteristics in terms of only one free parameter, 
namely the star formation rate of stars more massive than 5~$\msun$. 
Using the same scalings as in Scoville et al. 1997\nocite{SYB97}, but setting 
the minimum mass for the star forming rate to 5$\msun$, we obtain 
$\mdot (M \ge 5\msun) \approx 32 \msunyr$, which yields  
$\snrate \approx 1.5$\,yr$^{-1}$. 
Such a supernova rate would produce a non-thermal (synchrotron) luminosity of 
$\lsyn \approx 1.5\EE{23}$\whz at 1.4 GHz, which is  
about a 30\% lower than the observed value.
The star formation rate should be $\mdot (M \ge 5\msun) \gsim 40 \msunyr$ 
to be within 3$\sigma$ of the observed extended radio emission at 1.4 GHz
(3$\sigma \approx 0.4\EE{23}$\whz), and would
imply $\snrate \approx 2.0$\,yr$^{-1}$ \cite{Cond92}.
In turn, this would imply a $\lfir$ for component N of 
$7.7\times 10^{11} L_\odot$. Since
the total $\lfir$ for Mrk~273 is about $1.2\EE{12}~\lsun$, 
and given that DS98 do not provide quantitative estimates of the uncertainty 
for $\lfir$ of N, the above value for $\lfir$ is acceptable.
Therefore, the observed extended radio emission is easily explained within
those models and supports a scenario where it is due to relativistic electrons 
that have diffused away from SNR shocks.
Given the existing uncertainties, we estimate that a supernova
rate of $\snrate = 1.5$\,yr$^{-1}$ is likely to apply for Mrk~273, 
within an uncertainty of a factor of 2. 
A radio supernova would thus explode in the northern component 
of Mrk~273 approximately every 
eight months, and several supernovae would have been exploded between
the observations of CT00 on 1999, and ours in 2004.

\subsection{The South-Eastern Component}
It has been suggested that component SE is a
background source unrelated to Mrk~273, 
based on the lack of an NIR counterpart to component SE
\cite{Knap97}.
However, as already pointed out by CT00, the chances of having a
background source of $\sim 40$ mJy 
at an angular distance of 0.8 arcsec from component N are less
than $5\EE{-7}$. 
The 1.4 GHz images do not clarify whether the source is core-jet,
hence AGN driven, or a compact starburst, and observations
in other bands do not provide a unique interpretation.
At 5 GHz, the SE component is resolved in an arc-shaped radio emission
resembling a core twin-jet source.
The most striking peculiarity of the SE component  is its steep spectral index.
The integrated value is $\alpha\simeq 1.4$ with values ranging from 0.9 to
1.6 across the source. At the full resolution of the EVN observations (5 mas)
the source is completely resolved out, thus confirming the absence of any high
brightness radio feature and arguing against component SE being an AGN. 
In particular, given a $5\sigma$ limit of 0.13\,mJy,  we 
set an upper limit of $4\times 10^{20}$ WHz$^{-1}$ for the 5 GHz radio
luminosity of any AGN in this region.
One possibility could be that SE hosts also a compact starburst.
However, in that case we should expect the existence of diffuse emission 
in an extended region.
Instead, both the 1.4 GHz image of CT00 and our 5 GHz image 
(see Fig.~\ref{fig,SE}) show component SE to be rather compact, yet 
morphologically very much different.  
This frequency dependent morphology should not exist 
if we were dealing with a starburst scenario.
Hence, despite the milliarcsecond
resolution and high sensitivity of our new images, the physical nature of
component SE remains elusive, and unveiling it will have to await 
until more sensitive, multi-band observations are carried out.

\section{Summary}
We have presented the results from the most sensitive high-resolution
observations of the nuclear region in the ULIRG Mrk~273 ever made. 
Previous observations supported the scenario of a weak AGN 
and an extremely
luminous starburst coexisting in the nucleus of Mrk~273. We tested this
hypothesis using our new VLBI data at 1.6 and 5 GHz together with the 1.4 GHz
images published by CT00.

The main results of our analysis can be summarized as follows:
\begin{enumerate}
\item
Component N1, previously pinpointed as the radio counterpart of the AGN,
displays a steep spectral index between 1.4 and 5 GHz 
($\alpha = 1.2 \pm 0.1; S_\nu\, \propto\, \nu^{-\alpha}$).  
Given the steep radio 
spectrum is highly unlikely that N1 is  hosting the AGN,  and it 
rather suggests that the compact
non-thermal radio emission is produced by an individual RSN (which would
then be three to four times more luminous than SN~1988Z), or by a
combination of unresolved emission from several nested SNR and/or RSN.
\item
Component N2 displays both a rather complex and intriguing morphology and
spectral properties. The 5 GHz radio image is indicative 
of several compact features embedded in some diffuse
emission extended over a region of about 30 pc. The integrated spectral index
of this region is flat ($\alpha = 0.15 \pm 0.10$), which can be interpreted
as due to the superposition of several unresolved components, e.g., RSNe, 
SNR, or both,  whose radio emission peaks 
at different frequencies and is partially free-free absorbed.
Is it also possible that one of the compact components in this
region is the radio counterpart of the AGN.

\item
The radio morphology of component N as a whole supports the hypothesis that 
this is indeed the site of an extremely active star forming region. 
Following Scoville et al. (1997)\nocite{SYB97}, and using the constraints on 
the IR luminosity from the inner 120 pc central region around the nucleus
\cite{DS98}, a starburst lasting for $t_B=5\EE{7}$~yr, with
a Miller-Scalo initial mass function characterized by 
$m_l \approx 3.0 \msun$, $m_u \approx 30\msun$, and 
a sustained star formation rate of $\mdot \approx 39 \msunyr$ can easily
yield the observed FIR and radio emission.
Such a starburst would result into an estimated supernova rate for Mrk~273
of $\snrate \approx 1.5$\,yr$^{-1}$, and thus several supernovae
would have exploded between the observations of CT00 on 1999, and ours in 2004.

\item
The extended radio emission in component N has the typical spectral index
of non-thermal, optically thin emission ($\alpha = 0.8 \pm 0.1$), and 
luminosity ($L_{1.4 \rm GHz} = (2.2 \pm 0.1)\times 10^{23}$ WHz$^{-1}$) 
consistent with being
produced by relativistic electrons diffused away from supernova remnants 
in a recent outburst.

\item
Finally, the origin and interpretation of the radio emission from Mrk 273
SE  remains unclear. It has a very steep spectral index 
($\alpha = 1.4 \pm 0.2$), with no compact and/or  flat spectrum feature.
This would favour a starburst origin for the radio emission, 
although the lack of NIR emission poses some problems.
\end{enumerate}

\section*{Acknowledgments}
We thank Chris Carilli for having kindly provided the processed VLBA images 
at 1.4 GHz of Mrk~273, and the staff at JIVE for the efforts spent during the
correlation and pipeline of these data.
This work has benefited from research funding from the European Community's
Sixth Framework Programme (FP6).
The European VLBI Network is a joint facility of European, Chinese,
South African and other astronomy institutes funded by their national
research councils. MERLIN is a national facility operated by the University of
Manchester on behalf of PPARC.
This research has made use of the NASA/IPAC Extragalactic
Database (NED) which is operated by the Jet Propulsion
Laboratory, California Institute of Technology, under contract with the
National Aeronautics and Space Administration.


\begin{thebibliography}{}

\bibitem[Carilli \& Taylor 2000]{CT00}
Carilli, C.L., Taylor, G.B., 2000, ApJ, 532, L95 (CT00)

\bibitem[Cole et al. 1999]{Cole99}
Cole, G.H.J., Pedlar, A., Holloway, A.J., Mundell, C.G., 1999, 
MNRAS, 310, 1033

\bibitem[Condon 1992]{Cond92}
Condon, J.J., 1992, ARAA, 30, 575

\bibitem[Downes \& Solomon 1998]{DS98}
Downes, D., Solomon, P.M., 1998, ApJ, 507, 615

\bibitem[Falcke et al. 2000]{Falc00}
Falcke, H., Nagar, N.M., Wilson, A.S., Ulvestad, J.S., 2000, ApJ, 542, 197

\bibitem[Genzel et al. 1998]{Genz98}
Genzel, R. et al. 1998, ApJ, 498, 579

\bibitem[Kl\"ockner \& Baan 2004]{KB04}
Kl\"ockner, H.-R., \& Baan, W.A., 2004, A\&A, 419, 887

\bibitem[Knapen et al. 1997]{Knap97}
Knapen, J.H., Laine, S., Yates, J.A., Robinson, A., Richards, A.M.S.,
Doyon, R., Nadeau, D., 1997, ApJ, 490, L29


\bibitem[Majewski et al. 1993]{Maje93}
Majewski, S.R., Hereld, M., Koo, D.C., Illingworth, G.D., Heckman, T.M.,
1993, ApJ, 402, 125

\bibitem[Mazzarella \& Boroson 1993]{MB93}
Mazzarella, J.M., Boroson, T.A., 1993, ApJS, 85, 27

\bibitem[Middelberg et al. 2004]{Midd04}
Middelberg, E., et al. 2004, A\&A, 417, 925


\bibitem[Nagar, Wilson \& Falcke 2001]{NWF01}
Nagar, N.M., Wilson, A.S., Falcke, H., 2001, ApJ, 559, L87

\bibitem[O'Dea 1998]{Odea98}
O'Dea, C.P., 1998, PASP, 110, 493

\bibitem[Sanders et al. 1988]{Sand88} 
 Sanders, D.B., Soifer, B.T., Elias, J.H., Madore, B.F., Matthews, K., 
 Neugebauer, G., Scoville, N.Z., 1988, ApJ, 325, 74

\bibitem[Scoville, Yun \& Bryant 1997]{SYB97}
Scoville, N.Z., Yun, M.S., Bryant, P.M., 1997, ApJ, 484, 702

\bibitem[Smith et al. 1998]{Smit98}
Smith, H.E., Lonsdale, C.J., Lonsdale, C.J., Diamond, P.J., 1998, ApJ, 493, L17

\bibitem[Ulvestad \& Ho 2001]{UH01}
Ulvestad, J.S., Ho, L.C., 2001, ApJ, 558, 561

\bibitem[Veilleux, Sanders \& Kim 1999]{VSK99}
Veilleux, S., Sanders, D.B., Kim, D.C., 1999, ApJ, 522, 113

\bibitem[Weiler et al. 2002]{Weil02}
Weiler K.~W., Panagia N., Montes M.~J., Sramek R.~A., 2002, ARA\&A, 40, 387

\bibitem[Xia et al. 2002]{Xia02}
Xia, X.Y., Xue, S.J., Mao, S., Boller, Th., Deng, Z.G., Wu, H.,
2002, ApJ, 564, 196

\bibitem[Yates et al . 2000]{Yate00}
Yates, J.A., Richards, A.M.S., Wright, M.M., Collett, J.L., Gray, M.D., 
Field, D., Cohen, R.J., 2000, MNRAS, 317, 28

\end{thebibliography}
\end{document}